\documentclass[%
aps,
amsmath,amssymb,
reprint,
superscriptaddress
]{revtex4-2}

\usepackage{graphicx}                           % optional for graphics
\usepackage{amsfonts, amsmath, amssymb, amsbsy} % essential for equations
\usepackage{siunitx}
\usepackage{url}                                % optional for internet links
\usepackage{textcomp}                           % essential to use the upright \mu-symbol
\usepackage{lipsum}
\usepackage{hyperref}
\usepackage{cleveref}
\usepackage{mathtools}
\usepackage[normalem]{ulem}
\usepackage{float}
\usepackage{framed}   % for frame around images

\hypersetup{
	breaklinks=true,
	colorlinks=true,
	linkcolor=[RGB]{40,42,143},
	filecolor=[RGB]{40,42,143},
	citecolor=[RGB]{40,42,143},      
	urlcolor =[RGB]{40,42,143},
}

\newcommand{\bx}  {\boldsymbol{x}}

\begin{document}
	%% the square bracket argument will send term to running head in
	%% preprint, or running foot in reprint style.
	
	\title{Polydisperse versus monodisperse microbubbles: A simulation study for contrast-enhanced ultrasound imaging}
	
	\author{A.~Matalliotakis}	
	\affiliation{Department of Imaging Physics, Faculty of Applied Sciences, Delft University of Technology, 2628 CJ Delft, the Netherlands}%Lines break automatically or can be forced with \\
	
	\author{M.D.~Verweij}
	\affiliation{Department of Imaging Physics, Faculty of Applied Sciences, Delft University of Technology, 2628 CJ Delft, the Netherlands}%Lines break automatically or can be forced with \\
	\affiliation{Department of Biomedical Engineering, Erasmus University Medical Center, 3000 CA Rotterdam, the Netherlands}
	\email{Corresponding author: M.D.Verweij@tudelft.nl}
	%% For preprint only,
	%optional, if you want want this message to appear in upper right corner of title page
	%\preprint{A.~Matalliotakis, JASA}			%===================================================================================================
	% ABSTRACT =========================================================================================
	%===================================================================================================
	\begin{abstract}
		\textit{Objective}: Contrast-enhanced ultrasound (CEUS) presents distinct advantages in diagnostic echography. Utilizing microbubbles (MBs) as conventional contrast agents enhances vascular visualization and organ perfusion, facilitating real-time, non-invasive procedures. There is a current tendency to replace the traditional polydisperse MBs by novel monodisperse formulations in an attempt to optimize contrast enhancement and guarantee consistent behavior and reliable imaging outcomes. This study investigates the contrast enhancement achieved by monodisperse MBs of different sizes, and their influence on nonlinear imaging artifacts observed in traditional CEUS. 
		
		\noindent \textit{Methods}: To explore the differences between monodisperse and polydisperse populations without excessive experimentation, numerical simulations are employed for delivering precise, objective and expeditious results. The Iterative Nonlinear Contrast Source (INCS) method has previously demonstrated its efficacy in simulating ultrasound propagation in large populations in which each bubble has individual properties and several orders of multiple scattering are significant. Therefore, this method is employed to realistically simulate both monodisperse and polydisperse MBs.
		
		\noindent \textit{Results}: Our findings in CEUS imaging indicate that scattering from resonant monodisperse microbubbles is 11.8 dB stronger than scattering from the polydisperse population. Furthermore, the amplitude of nonlinear imaging artifacts downstream of the monodisperse population is 19.4 dB stronger compared to polydisperse suspension.
		
		\noindent \textit{Conclusion}: Investigating the impact of multiple scattering on polydisperse populations compared to various monodisperse suspensions reveals that monodisperse MBs are more effective contrast agents, especially when on resonance. Despite the strong signal to noise ratio of monodisperse populations, the imaging artifacts due to nonlinear wave propagation are also enhanced, resulting in more missclassification of MBs as tissue. 
		
	\end{abstract}
	%Use showkeys class option if keyword display desired
	%===================================================================================================
	
	\maketitle
	
	%  End of title page for Preprint option --------------------------------- %
	%
	%=================================================================================================
	\section{Introduction} \label{sec:Poly_Introduction}
	%=================================================================================================
	%
	Achieving superior deep tissue imaging of blood vessels with ultrasound remains a challenge in medical diagnostics. Contrast-enhanced imaging, particularly using MBs, has emerged as a promising solution \cite{Averkiou2020, Versluis2020}. These gas-filled microspheres, stabilized with a lipid or protein shell, enhance blood contrast for improved organ and lesion visualization. MBs, characterized by small size, biocompatibility, and vascular navigability, resonate in the ultrasound frequency range (1-10 MHz). Their efficient sound scattering in both fundamental and harmonic modes, driven by substantial acoustic impedance difference with surroundings and highly nonlinear oscillatory behavior \cite{deJong2002, Marmottant2005}, enhances image quality. As ultrasound waves propagate through a resonant MB suspension, they undergo nonlinear distortion due to nonlinear MB scattering influenced by size, shell characteristics, ultrasound pressure and frequency \cite{Emmer2009, Sojahrood2015, Sojahrood2023}. Because of these properties, MBs are also efficient contrast agents in various applications besides CEUS, such as ultrasound localization microscopy \cite{Errico2015}.
	
	As a drawback, wave distortion extends beyond a MB suspension and this leads to the misidentification of tissues as MBs, diminishing the specificity of CEUS imaging \cite{Tang2006}. Narrowing of the size distribution of the MB population might be a way to provide improved acoustic scattering, reduce imaging artifacts and enhance scattering homogeneity. Historically, polydisperse MBs with varying size distributions (typical radii 0.5 to 15 $\mu$m) have been standard in ultrasound contrast imaging \cite{Lindner2004,Frinking2020}. Recent technological breakthroughs have introduced the possibility of using monodisperse, i.e. uniformly sized, MBs \cite{Segers2016}. Studies highlight the superiority of monodisperse MBs \cite{Helbert2020}, offering enhanced predictability, improved acoustic performance, and clearer imaging signals \cite{Segers2018, vanElburg2023}.  Nevertheless, we think that it is important to shed more light on the effect of monodisperse MBs as contrast agents for deep vessel imaging, especially on the generation of clearer echoes and reducing imaging artifacts.
	
	The use of computational tools is an efficient way to perform comprehensive investigations without performing extensive measurements. Initially, studies focused on the collective behavior of bubbly media for marine applications \cite{Foldy1945,Foldy1947}. Effective medium theory facilitated 1D computational studies for both monodisperse \cite{Stride2005,Hibbs2007} and polydisperse \cite{Ando2011,Ovenden2017} MB suspensions in medical ultrasound, including high intensity focused ultrasound (HIFU) \cite{Vanhille2019}. Previous models successfully captured nonlinear ultrasound propagation through uniform MB distributions in two dimensions using iterative schemes \cite{Pinton2009, Joshi2017}. Challenges arise when coupling the nonlinear dynamics of multiple MBs in 3D realistic simulations, due to the complexity of the coupled Rayleigh-Plesset equation \cite{Haghi2019}. Another difficulty shows up when the number of polydisperse MBs is small and the use of averaged quantities becomes questionable. Various computational methods have been explored to understand the dynamics between polydisperse and monodisperse MB populations. Among these, the INCS method has demonstrated efficacy in simulating bubble cloud behavior in a three-dimensional domain, enabling the generation and comparison of echoes produced by dense monodisperse MB populations, considering multiple scattering \cite{Bubble_Cloud}. This is crucial for optimizing contrast-enhanced ultrasound (CEUS) applications and reducing the need for excessive experimentation. 
	
	The aim of this numerical study is to investigate the efficacy of monodisperse and polydisperse populations when used as contrast agents for deep tissue imaging. More precisely, this article discusses the extension of INCS method to simulate the behaviour of a population of polydisperse scatterers. Furthermore, the effectiveness of the extended INCS method is illustrated through simulating the multiple scattering occurring inside a population of polydisperse MBs, each with individual properties represented by its own Marmottant model \cite{Marmottant2005}. INCS is based on an iterative scheme for computing  the scattered acoustic signals \cite{Koos_Thesis, Huijssen2010}. Numerically, the accuracy of the final result is improved after each iteration. In a physical sense, each iteration adds an extra order of multiple scattering corresponding to an additional path of wave propagation.
	
	First, in Section~\ref{sec:Poly_INCS}, the fundamental theory behind the INCS method is explained, followed by its extension with the introduction of polydisperse point scatterers. In Section~\ref{sec:Poly_CompConfig}, the configurations for the numerical experiments are discussed. Next, in Section \ref{sec:Poly_Numerical_Results} the results from the numerical simulations for each different test case are presented. Concluding remarks are given in Section.~\ref{sec:Poly_Conclusions}.
	%=================================================================================================
	\section{Inclusion of a polydisperse MB population}\label{sec:Poly_INCS}
	%=================================================================================================
	%
	%========================================================
	\subsection{Linear Field}
	%========================================================
	The linear pressure field generated by an external source in a linear, lossless, homogeneous acoustic medium is described by the wave equation
	\begin{equation}
		c^{-2}_{0}\frac{\partial^{2} p(\bx,t)}{\partial t^2}-\nabla^2 p(\bx,t)= S_\mathrm{pr}(\bx,t).
		\label{eq:LosslessLinWestervelt}
	\end{equation}
	Here, $\bx$ [m] is the Cartesian position vector, and $t$ [s] is the time.  The symbol $p(\bx,t)$ [Pa] indicates the acoustic pressure, $c_{0}=1/\sqrt{\rho_{0}{\kappa_{0}}}$ [m/s] is the small signal sound speed in the background medium, where $\rho_{0}$ [kg$\cdot \mathrm{m}^{-3}$] is the mass density and ${\kappa_{0}}$ [$\mathrm{Pa}^{-1}$] is the compressibility. The Laplacian operator ${\nabla}^2$ generates the sum of the second order spatial derivatives. The acoustic field is generated by the primary source term $S_\mathrm{pr}$, which can for example describe a jump condition for either the velocity or the pressure. These jump conditions can be used to represent a source with a plane aperture, e.g. a phased array transducer.
	
	%========================================================
	\subsection{Nonlinear field due to contrast agents}
	%========================================================
	In medical ultrasound, nonlinearities arising from contrast media can have a significant impact on the propagation of the acoustic signals. To incorporate any phenomena that affect the pressure field, it is possible to extend Eq.~(\ref{eq:LosslessLinWestervelt}) with a contrast source term $S_\mathrm{cs}$
	\begin{equation}
		{c^{-2}_{0}}\frac{\partial^{2} p}{\partial t^2}-\nabla^{2}p = S_\mathrm{pr} + S_\mathrm{cs}(p). 
		\label{eq:LosslessContrast}
	\end{equation}
	With this approach, multiple contrast sources can be accommodated that represent global nonlinear effects \cite{Koos_Thesis,Huijssen2010}, attenuation \cite{Libe_Thesis,Libe2009}, inhomogeneous medium properties \cite{Libe2011}, or local nonlinear effects \cite{LocalNL2023}. In contrast-enhanced imaging, the nonlinear oscillatory behavior of the MBs influences the pressure field. To include the contribution of a population of $N$ MBs, each will be described as a point scatterer and the source term will be written as \cite{Bubble_Cloud}
	\begin{align}
		S_\mathrm{cs}(\bx,t)
		&= \sum_{i=1}^N S_{\mathrm{MB}_i} \nonumber\\
		&= \rho_0\sum_{i=1}^N\frac{d^2 V^{(i)}(\bx_{\mathrm{sc}}^{(i)},t)}{d t^2}\,\delta(\bx-\bx_{\mathrm{sc}}^{(i)}),
		\label{eq:Distribution_PS_SourceTerm}
	\end{align} 
	where $V^{(i)}$ is the volume of the $i$th MB, $\bx^{(i)}_{\mathrm{sc}}$ is the position vector of its center and $\delta$ is the Dirac delta distribution.  Each scatterer's volume depends on the bubble radius $R$ as a function of time, which in our case will be calculated by solving the Marmottant equation \cite{Marmottant2005,Bubble_Cloud}.
	
	In the case of a population of monodisperse MBs, the rest radius $R_0$ is the same for all the scatterers, whereas for a polydisperse distribution, each scatterer has its own rest radius $R_0^{(i)}$.
	%
	%
	%=================================================================================================
	\section{Configurations used in the simulations}\label{sec:Poly_CompConfig}
	%=================================================================================================
	%
	\begin{figure}[t!]
		\centering
		\includegraphics{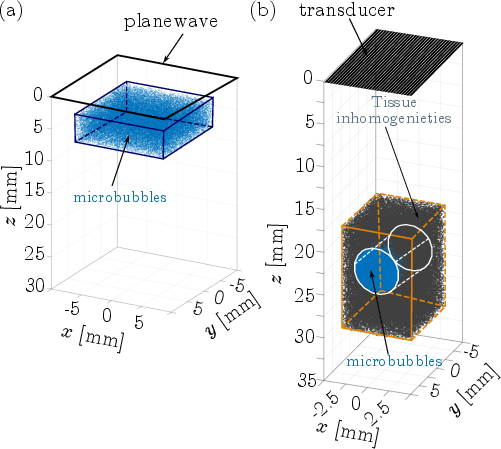}% Here is how to import EPS art
		\caption{Configurations used in the INCS simulations. (a) Computational domain containing a suspension of 3.5 $\times 10^4\;\mathrm{ml}^{-1}$ (blue) monodisperse MB with $3.2\;\mu\mathrm{m}$ rest radius, embedded in water. (b) Computational domain incorporating $7\times 10^5\;\mathrm{ml}^{-1}$ tissue-mimicking linear scatterers (grey) surrounding a suspension of $5\times 10^5\;\mathrm{ml}^{-1}$ monodisperse MBs with $1.4\;\mu\mathrm{m}$ rest radius (blue). }
		\label{fig:ContrastDomain}
	\end{figure}
	\subsection{Simulation of pressure fields} \label{subsubsec:Poly_ContrastDomain}
	\subsubsection{Incident field and contrast domain}
	In this study, we will consider the computational domain and the domain for the contrast media as depicted in Fig.~\ref{fig:ContrastDomain}(a). This configuration is used in Secs.~\ref{subsec:Poly_Validation} and \ref{subsec:Poly_PW_Results} for the INCS validation and the comparison between different populations,  respectively. The computational domain has dimensions $X\times Y\times Z=20\;\mathrm{mm}\times 20\;\mathrm{mm}\times 30\;\mathrm{mm}$ is used. The scatterers are placed in a domain with dimensions $X\times Y\times Z=15\;\mathrm{mm}\times 15\;\mathrm{mm}\times 4.444\;\mathrm{mm}$, resulting in a $1\;\mathrm{ml}$ volume. These configuration choices are made to simplify the comparison between polydisperse and monodisperse populations.	
	
	The incident pressure field is a plane wave being generated at $z=0$ and propagating in the positive $z$-direction. A plane wave is used to let all the scatterers experience the same incident pressure. The temporal signature of the incident pressure is
	\begin{equation}
		p(t)=P_0\,\mathrm{exp}\left[ -\left(\frac{t-T_d}{T_w/2}\right)^2\right]\mathrm{sin}[2\pi f_0 (t-T_d)],
		\label{eq:Time_Signature}
	\end{equation}
	where $T_w=3/f_0$ is the width and $T_d=6/f_0$ is the delay of a Gaussian envelope with a duration of $12/f_0$, where $f_0 = 1\;\mathrm{MHz}$ is the center frequency. The peak pressure is $P_0=200\;\mathrm{kPa}$. The scatterers will be embedded in water with a density of $\rho = 1060\ \mathrm{kg/m^3}$ and a speed of sound of $c_0 = 1482\;\mathrm{m/s}$. In the considered situations, water has negligible losses and nonlinear effects will be hardly noticeable. Therefore, we assume that the embedding medium is lossless and linear. A sampling frequency of 18 MHz was used as the basis for the discretization of the spatiotemporal domain \cite{Bubble_Cloud}.
	\bigskip
	\subsubsection{Configuration for validation}
	\label{subsubsec:config_validation}
	To validate INCS we compare our results by those following from effective medium theory. The analytical expressions that describe the effective behavior of a population of isotropic linear scatterers (LSs) are derived from Foldy \cite{Foldy1945,Foldy1947}. The same approach has been used in a previous publication for a monodisperse population of scatterers \cite{Bubble_Cloud}, but here we will consider a polydisperse population. For the INCS implementation, we assume that the contrast source term for each LS is given by
	\begin{equation}
		S_\mathrm{sc}(\bx,t)= -f(R_0)\,V_0\,\frac{\rho_0}{\rho_1 c_1^2}\,
		\frac{\partial^2 p(\bx_\mathrm{sc},t)}{\partial t^2}\,\delta(\bx-\bx_\mathrm{sc}),
		\label{eq:SC_monopole_time_domain_approx}
	\end{equation}
	where $R_0$ is rest radius, $V_0$ is its initial volume, and $f(R_0)$ is the polydispersity coefficient given by
	\begin{equation}
		f(R_0) = \frac{k}{(R_0/R_\mathrm{0,ref})^\gamma}.
		\label{eq:polydispersitycoeff}
	\end{equation}
	$k$ is a constant to adjust the scattering strength if necessary and $\gamma$ is the polydispersity scale parameter to control the scattering distribution of the population. 
	
	In the case of a plane wave excitation as in Eq.~(\ref{eq:Time_Signature}), the scattered pressure is given by
	\begin{align}
		p_\mathrm{sc}(\bx,\omega)
		&= f(R_0)\,V_0\,\frac{\rho_0}{\rho_1 c_1^2}\omega^2\;\frac{p(\omega)}{4\pi r}e^{-ikr} \nonumber\\
		&= g(R_0,\omega)\,\frac{p(\omega)e^{-ikr}}{r},
		\label{eq:Scattering_strength}
	\end{align}
	where $g(R_0,\omega)$ is the scattering strength of an individual LS, and r is the distance from the scatterer.  We follow this approach in order to match the variables as defined previously in Foldy \cite{Foldy1945}.
	
	In this study we assume $k=0.6$, $\gamma=2$ and $R_{0,ref}=1\ \mu\mathrm{m}$, resulting in a scattering strength that is a linear function of the (fictitious) radius of a scatterer. Thus, there will not be extremely large differences in the scattering strength in the polydisperse population.
	
	For the polydisperse populations considered in this paper, the density of the microbubbles varies with the rest radius $R_0$ according to the gamma distribution
	\begin{equation}
		n(R_0)=\frac{N}{V}\,\frac{1}{b^\alpha\Gamma(\alpha)}\,
		R_0^{\alpha-1}e^{-R_0/b},
		\label{eq:gamma_distr}
	\end{equation}
	where $N$ is the total number of scatterers, and $V$ is the volume in which the homogeneous population resides. Furthermore, $\alpha$ and $b$ are the scale and shape parameters, and $\Gamma$ is the gamma function \cite{Gamma2013}. In our case we take $\alpha=2.24$ and $b=1.23\,\mu\mathrm{m}$. The total density of the microbubbles with radii between $R_\mathrm{0,min}$ and $R_\mathrm{0,max}$ is given by
	\begin{equation}
		n_\mathrm{tot} = \int_{R_\mathrm{0,min}}^{R_\mathrm{0,max}}\; n(R_0)\; dR_0,
		\label{eq:Int_n}
	\end{equation}
	where $R_\mathrm{0,min}$ and $R_\mathrm{0,max}$ are the minimum and maximum values considered to be present within the polydisperse distribution. For the given parameters $a$ and $b$, virtually all microbubbles are taken into account if we take $R_\mathrm{0,min}=0.5\;\mu\mathrm{m}$ and $R_\mathrm{0,max}=15\; \mu\mathrm{m}$.
	\subsubsection{Types of monodisperse and polydisperse suspensions}\label{subsubsec:Poly_PW_Config}
	To make a comparison between the efficiency of a population of monodisperse and polydisperse MBs, we take into account four distinct populations:
	\begin{enumerate}
		\item A monodisperse population of MBs with a rest radius $R_0=4\;\mu\mathrm{m}$ and a resonance frequency $f_\mathrm{res}=0.8\;\mathrm{MHz}$ (below the center excitation frequency);
		\item A monodisperse population of MBs with a rest radius $R_0=3.2\;\mu\mathrm{m}$ and a resonance frequency $f_\mathrm{res}=1\;\mathrm{MHz}$ (at the center excitation frequency);
		\item A monodisperse population of MBs with a rest radius $R_0=1\;\mu\mathrm{m}$ and a resonance frequency $f_\mathrm{res}=3.9\;\mathrm{MHz}$ (above the center excitation frequency);
		\item A polydisperse population of MBs with a rest radius between $R_\mathrm{0,min}=0.5\;\mu\mathrm{m}$ to $R_\mathrm{0,max}=15\; \mu\mathrm{m}$, distributed as described in Section \ref{subsubsec:config_validation}, corresponding to a resonance frequency between $f_\mathrm{res}=0.3\;\mathrm{MHz}$ and $10\;\mathrm{MHz}$ (a number of MBs will be near the resonance frequency, others will be above of below resonance)
	\end{enumerate}
	
	In our simulations, we use high driving pressures to activate the nonlinear oscillatory behaviour of the MBs and therefore the contribution of the shell stiffness becomes unimportant. As a result, the resonance frequency of the MBs shifts towards the resonance of an uncoated bubble \cite{Versluis2020}. Thus, we have approximated the resonance frequency by the eigenfrequency \cite{Overvelde2010} 
	\begin{equation}
		f_\mathrm{res} = \frac{1}{2\pi R_0}\sqrt{\frac{1}{\rho_0}\left[3 \gamma  \cdot P_\mathrm{amb} + (3\gamma-1)\frac{2\sigma_w}{R_0} \right]},
		\label{eq:f_uncoated}
	\end{equation}
	where $R_0$ is the initial radius of the MB, $\gamma=1.07$ is the polytropic exponent of the gas encapsulated in the bubble, and $P_\mathrm{amb}=101.3 $ kPa is the static ambient pressure. The center excitation frequency $f_0=1\;\mathrm{MHz}$ corresponds to a resonance frequency of an uncoated MB of radius $R_0=3.2\;\mu\mathrm{m}$.
	
\begin{figure}[t!]
		\centering
		\includegraphics{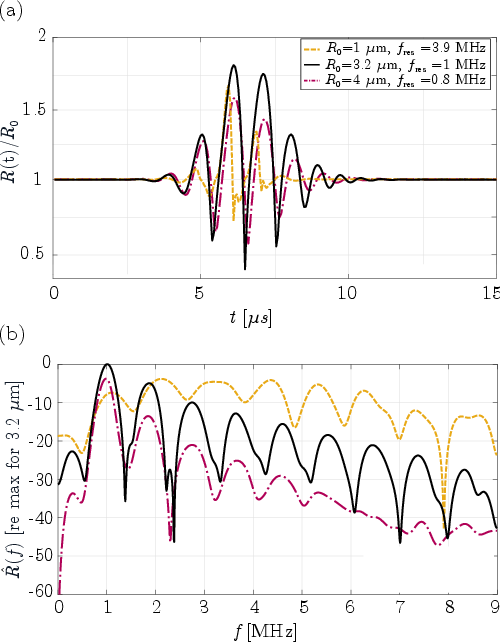}
		\caption{(a) Temporal radial responses $R(t)$ of MBs with rest radii of $1\;\mu\mathrm{m}$, $3.2\;\mu\mathrm{m}$, and $4\;\mu\mathrm{m}$, when excited by a 3 cycle pulse with 200 kPa peak pressure and 1 MHz center frequency. (b) The frequency spectra $\hat{R}(f)$ corresponding to the signals in (a). }		 
		\label{fig:Radius_Frequency}
	\end{figure}

	For solving the Marmottant equation \cite{Marmottant2005}, we further use the gas core viscosity $\mu=2\times 10^{-3}\;\mathrm{Pa}\cdot\mathrm{s}$, the effective surface tension $\sigma(R)=0.036\;\mathrm{N/m}$, the shell elasticity $\chi=0.4\;\mathrm{N/m}$, and the surface tension of the gas-water interface $\sigma_w=0.072\;\mathrm{N/m}$ \cite{Segers2018, Helbert2020}. The shell elasticity is given by $\kappa_{s}=1.5\times 10^{-9}\,\exp{(8\times 10^5 R_0)}$ \cite{Segers2018a}. Combined with the aforementioned, the oscillatory behavior and the frequency spectrum of a single MB when excited with a driving pressure $P_0=200\;\mathrm{kPa}$ and a center frequency $f_0=1\;\mathrm{MHz}$ is depicted in Fig.~\ref{fig:Radius_Frequency}.
	\subsection{Simulation of CEUS imaging} \label{subsec:MedicalApp}	
	To actually see the difference between monodisperse and polydisperse populations for contrast-enhanced imaging, it is necessary to visualize the reconstructed beamformed images from the scattered radio frequency (RF) data generated by a realistic configuration. To mimic tissue with an enclosed vessel, we distribute LSs surrounding a cylindrical population of MB, as depicted in Fig.~\ref{fig:ContrastDomain}(b). We need to take into account all the relevant phenomena that occur during the propagation of ultrasound through the populations of scatterers inside the water background medium. Based on this, the new nonlinear wave equation is given by 
	\begin{equation}
		\label{eq:LosslessContrast}
		c^{-2}_0\partial^{2}_t p-\nabla^2 p = S_\mathrm{pr} + S_\mathrm{MBs}(p) + S_\mathrm{LSs}(p) + S_\mathrm{nl}(p) + S_\mathrm{\mathcal{L}}(p),
	\end{equation}
	where $S_\mathrm{MBs}$ is the contrast source term for the MB population \cite{Bubble_Cloud}, $S_\mathrm{LSs}$ is the contrast source term for the LS population \cite{Bubble_Cloud}, $S_\mathrm{nl}$ and $S_\mathrm{\mathcal{L}}$ are the terms for global \cite{Bubble_Cloud} and local medium nonlinearities \cite{LocalNL2023}, respectively. Equation~(\ref{eq:LosslessContrast}) is solved iteratively using a Neumann scheme, as described in previous publications \cite{Bubble_Cloud,XwaveMBs2023}. 
	
	The incident pressure field is computed for a P4-1 probe (Verasonics, Washington, USA). Transducer elements have a height of $H_{\mathrm{el}} = 16\ \mathrm{mm}$, a width of $W_{\mathrm{el}}=0.245\ \mathrm{mm}$, a pitch of $D_{\mathrm{tr}}=0.295\ \mathrm{mm}$. The transmitted pulse is given by Eq.~(\ref{eq:Time_Signature}), with center frequency $f_0=2.5\;\mathrm{MHz}$ and a peak pressure at the transducer surface of $P_0=200\;\mathrm{kPa}$, to activate the nonlinear behavior of the monodisperse MBs. Next, the domain of the MB population is a cylinder with center $(x,\ y,\ z)=(0,\ 0,\ 22.5)$ mm, diameter of $5\;\mathrm{mm}$ and length of $10\;\mathrm{mm,}$ as illustrated in Fig.~\ref{fig:ContrastDomain}(b). This corresponds to a total volume of $0.2\;\mathrm{ml}$. The domain of LSs surrounding the MBs is a cube of $X\times Y\times Z=8\;\mathrm{mm}\times 10\;\mathrm{mm}\times 12\;\mathrm{mm}$, corresponding to a volume of $0.76\;\mathrm{ml}$ centered at $(x,\ y,\ z)=(0,\ 0,\ 24)$ mm, as depicted in Fig.~\ref{fig:ContrastDomain}(b). Furthermore, the background medium is water, with a coefficient of nonlinearity of $\beta=3.21$.
	
	To accurately solve the full nonlinear wave equation up to the second harmonic frequency ($h=2$)  of the incident pressure pulse, we need to have a Nyquist frequency of at least $F_\mathrm{nyq} = (h+1.5)f_0=3.5 f_0$. To also safely capture the higher harmonics of the MB scattering, we used $F_\mathrm{nyq}=5 f_0=12.5$ MHz. Thus, the sampling frequency, used for discretizing the spatiotemporal domain, is $F_\mathrm{s}=2F_\mathrm{nyq}=25\;\mathrm{MHz}$. Furthermore, we need at least $j=h+1=3$ iterations for an accurate prediction of the second harmonic \cite{Huijssen2010}. We take $j=10$ iterations to ensure that the relative root mean square error between successive iterations is below $10^{-6}$. This also implies that our simulations account for MB interactions up to ninth order multiple scattering \cite{Bubble_Cloud}.
	
	We compare CEUS imaging with two different microbubble populations:
	\begin{enumerate}
		\setcounter{enumi}{4}
		\item A resonant monodisperse population of MBs with a rest radius $R_0=1.4\;\mu\mathrm{m}$ and a resonance frequency $f_\mathrm{res}=2.5\;\mathrm{MHz}$ (at the center excitation frequency);
		\item A polydisperse population of MBs with a rest radius between $R_0=0.5\;\mu\mathrm{m}$ and $15\;\mu\mathrm{m}$, distributed as described in Section \ref{subsubsec:config_validation}, and a resonance frequency between $f_\mathrm{res}=0.3\;\mathrm{MHz}$ and $10\;\mathrm{MHz}$.
	\end{enumerate}
	Each LS has a scattering strength which can be computed through Eq.~(\ref{eq:SC_monopole_time_domain_approx}), for a polydispersity coefficient $f=1$.
	
	For the beamforming process, we use the MUST \cite{MUSTToolbox} toolbox after employing the amplitude modulation (AM) technique and a virtual point source formulation as described by Garcia et. al. \cite{DAS2021}. 
	%=================================================================================================
	\section{Numerical results}\label{sec:Poly_Numerical_Results}
	%=================================================================================================
	%
	\subsection{Comparison of INCS and effective medium theory}\label{subsec:Poly_Validation}
	\begin{figure}[b!]
		\centering
		\includegraphics{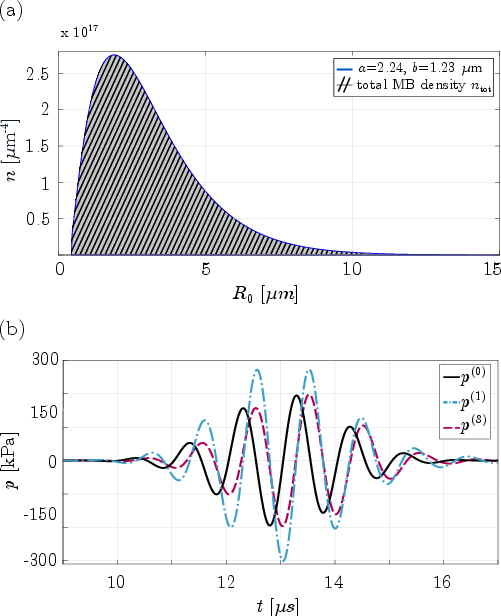}
		\caption{(a) Graph depicting the continuous gamma distribution with $\alpha=2.24$ and $b=1.23\;\mu\mathrm{m}$ (blue). For $R_\mathrm{0,min}=0.5\;\mu\mathrm{m}$ and $R_\mathrm{0,max}=15\; \mu\mathrm{m}$, the area below the curve equals to a concentration $n_\mathrm{tot}=10^6$ $\mathrm{ml}^{-1}$. (b) Comparison between the time signatures of the incident pressure pulse (p$^{(0)}$, black continuous line), and the total pressure pulse after $j=1$ (p$^{(1)}$, blue dotted line) and $j=8$ (p$^{(8)}$, magenta dashed line) iterations that is received by a point receiver located on the $z$ axis at $z=10.3\:\mathrm{mm}$. }		 
		\label{fig:Validation}
	\end{figure}
	\begin{figure*}[t!]
		\centering
		\includegraphics{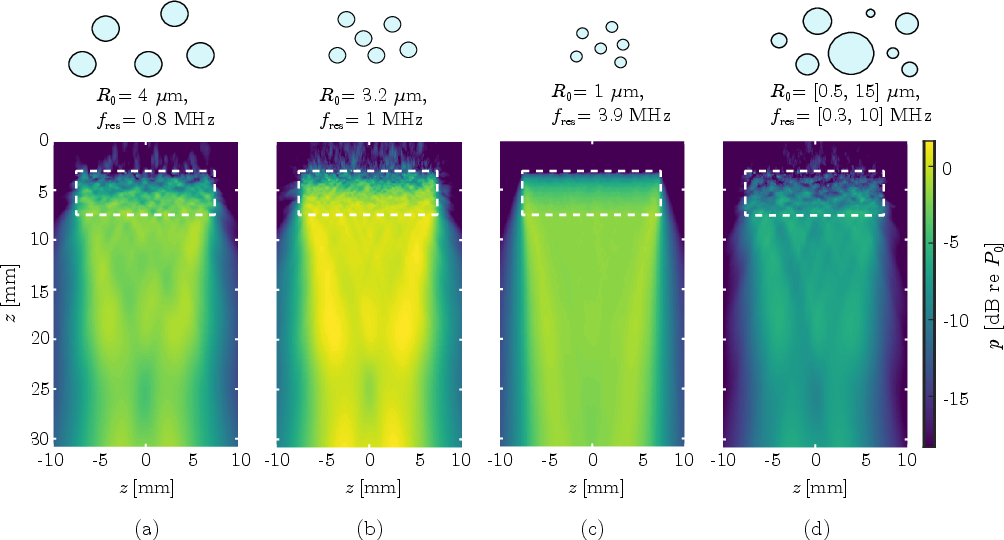}
		\caption{Peak of the total scattered pressure at $y=0$ mm for (a) a type 1 monodisperse distribution of MBs, (b) a type 2 monodisperse distribution of MBs, (c) a type 3 monodisperse distribution of MBs, and (d) a type 4 polydisperse distribution of MBs. In all four cases, the population is located inside the dashed white rectangle. }		 
		\label{fig:Comp_Scattered_Pressure}
	\end{figure*}
	In this section, we assume that there are $N=10^6$ microbubbles located in the $V=1\;\mathrm{ml}$ volume indicated in Fig.~\ref{fig:ContrastDomain}. The suspension has a type 4 polydisperse distribution, as described in Section \ref{subsubsec:Poly_PW_Config}. The total gas volume corresponds to $7.41\times 10^{-6}\;\mathrm{ml}$. It is assumed that the gas inside the bubbles is $C_4F_{10}$, with a density $\rho_1 = 10 \  \mathrm{kg}/\mathrm{m}^3$ and a speed of sound $c_1 = 100\ \mathrm{m}/\mathrm{s}$. As we want to perform a simplified comparison with effective medium theory, we do not take into account the resonance frequency and the nonlinear behavior of the microbubbles. Instead, we assume that each bubble can be described by its scattering behavior as describe in Eqs.~(\ref{eq:SC_monopole_time_domain_approx}) to (\ref{eq:Scattering_strength}). In other words, we are only interested on the scattered signal of each point scatterer. The maximum of the incident pressure $P_0 = 200\ \mathrm{kPa}$ will not affect the final result because we operate in the linear regime.

	According to Foldy's theory \cite{Foldy1945,Foldy1947}, the effect of a polydisperse population of scatterers is represented by replacing the wave number $k_0$ in the scattering domain by a corrected wave number $k$ according to
	\begin{equation}
		k^2=k_0^2 + 4\pi \int_{R_\mathrm{0,min}}^{R_\mathrm{0,max}} g(R_0,\omega)\;n(R_0)\;dR_0,
		\label{eq:k_foldy_poly}
	\end{equation}
	where $g(R_0,\omega)$ $[\mathrm{m}]$ is derived from Eq.~(\ref{eq:Scattering_strength}), and $n(R_0)$ is computed through Eq.~(\ref{eq:gamma_distr}. The shift in wavenumber corresponds to a shift in wave speed, and as a consequence, in a time shift of the wave that has traversed the scattering domain. In the case considered in this subsection, the integral amounts to $2.3\times10^{5}\;\mathrm{m}^{-2}$. This yields a wavespeed of $1375.5\;\mathrm{m/s}$ in the scattering domain, while the speed in the medium without scatterers is $1482\;\mathrm{m/s}$. Since the scattering domain has a length of $4.4444\;\mathrm{mm}$, the additional time delay caused by the scattering domain, as predicted by the theory of Foldy, is $\Delta t_\mathrm{Foldy}=0.228\;\mu\mathrm{s}$. We have also determined the time delay between the incident wave $p^{(0)}$ and the wave with all significant orders of scattering $p^{(8)}$ from Fig.~\ref{fig:Validation}(b), by looking at the shift in the zero crossings around $13\;\mu\mathrm{s}$. This is found to be $\Delta t_\mathrm{INCS}=0.232\;\mu\mathrm{s}$. Thus, the difference in time delay as predicted by the theory of Foldy and our method is only 1.75\%.
	
	Furthermore, since the wavenumber derived from Eq.~(\ref{eq:k_foldy_poly}) lacks an imaginary component in our specific case, according to Foldy's theory \cite{Foldy1945,Foldy1947}, the wave traversing the scattering domain is not subject to attenuation. As illustrated in Fig.~\ref{fig:Validation}(b), in our approach the later iterations correct the larger amplitudes observed in earlier iterations, and iteration $p^{(8)}$ has the same amplitude as the incident field $p^{(0)}$. This consistency in both time delay and wave amplitude across a scattering domain indicates a good quantitative agreement between our method and Foldy's effective medium theory in case of a polydisperse distribution of scatterers.  
	\subsection{Plane wave: monodisperse vs polydisperse}\label{subsec:Poly_PW_Results}
	\begin{figure*}[t!]
		\centering
		\includegraphics{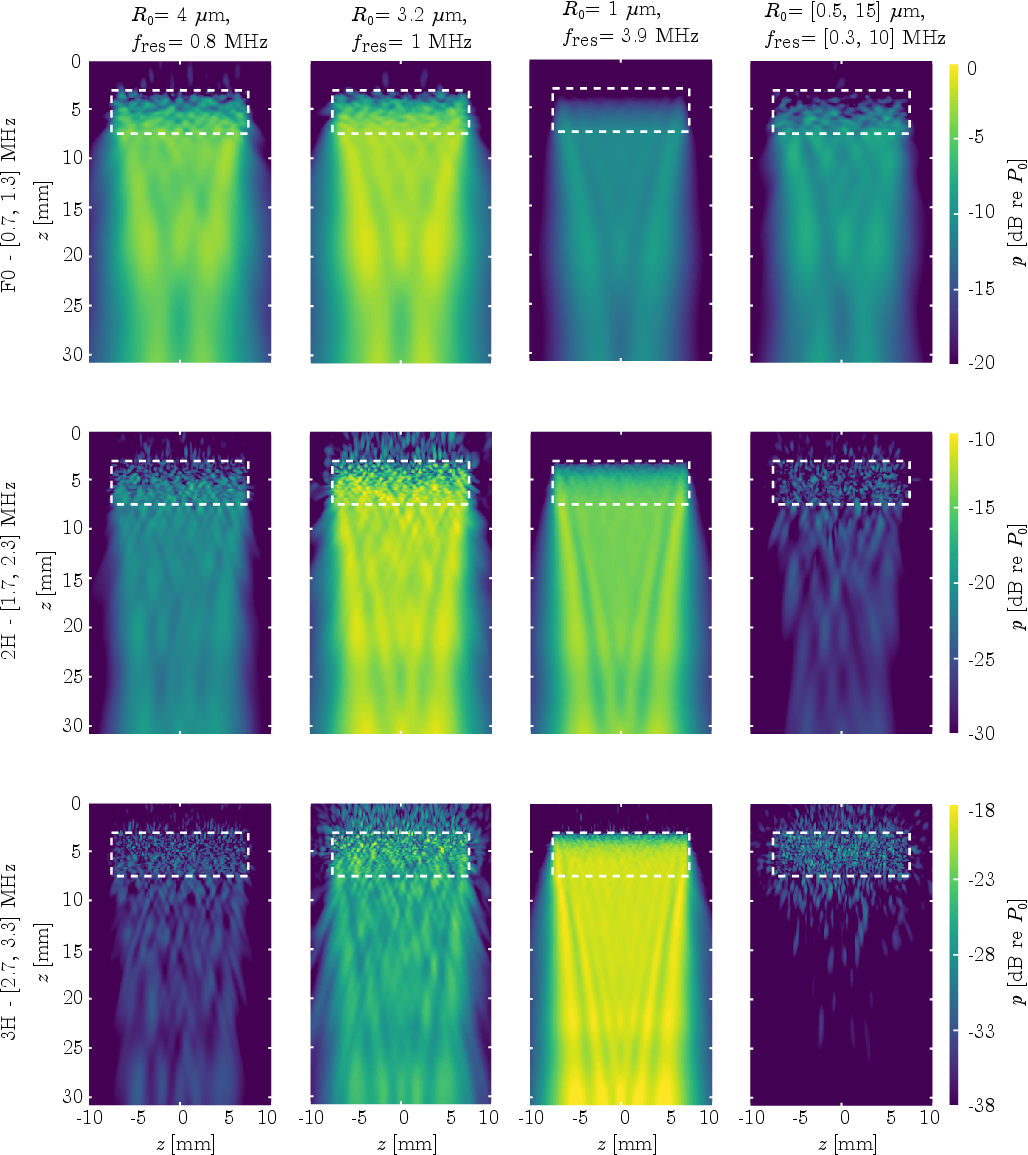}
		\caption{Peak of the scattered pressure at $y=0$ mm for: (first column) a type 1 monodisperse distribution of MBs, (second column) a type 2 monodisperse distribution of MBs, (third column) a type 3 monodisperse distribution of MBs, and (fourth column) a type 4 polydisperse distribution of MB. Each row corresponds to a specific frequency band: (first row) fundamental F0 [0.7, 1.3] MHz, (second row) second harmonic 2H [1.7, 2.3] MHz, and (third row) third harmonic 3H [2.7, 3.3] MHz. In all four cases, the population is located inside the dashed white rectangle.}		 
		\label{fig:Comp_Scattered_Pressure_Harm}
	\end{figure*}
	We continue with a comparison between four different populations of MBs, as mentioned in Sec.~\ref{subsubsec:Poly_PW_Config}. To start, our reference is the type 2 monodisperse resonant population, for which we use 35,000 MBs, resulting on a total gas volume of $4.6\times 10^{-6}\;\mathrm{ml}$. To achieve a fair comparison, the total gas volume concentration of the MB suspension should be the same in all the cases \cite{Bubble_Cloud}.  Therefore, the type 1 monodisperse population will contain 17,500 MBs, the type 3 monodisperse population will consist of $10^5$ MBs and the type 4 polydisperse population will count 20,000 MBs. The bubble populations are placed in the volume $V=1\;\mathrm{ml}$ as indicated in Fig.~\ref{fig:ContrastDomain}(a).
	
	\subsubsection{Scattered pressure field: Full spectrum}
	The scattered pressure field in each case is depicted in Fig.~\ref{fig:Comp_Scattered_Pressure}. At first sight, the scattered pressure generated from the resonant MBs (type 2, $R_0=3.2\;\mu\mathrm{m}$) is the strongest between all the cases with a peak pressure of $+1.1\;\mathrm{dB}$ relative to the peak incident pressure $P_0$. Next, the case below resonance (type 1, $R_0=4\;\mu\mathrm{m}$) follows with a relative peak amplitude of $-1.13\;\mathrm{dB}$. Although these MBs have a resonance frequency that is still close to the excitation frequency, their peak amplitude is significantly smaller than the resonant MBs. The third case with a relative peak amplitude of $-1.45\;\mathrm{dB}$ is the one above resonance (type 3, $R_0=1\;\mu\mathrm{m}$) and the last one is the polydisperse distribution (type 4) with a peak pressure of $-5.47\;\mathrm{dB}$. These results demonstrate that when the resonance frequency is closer to the excitation frequency then the scattered pressure field is stronger, with the scattering of the resonant contrast agents being the highest. Another observation is that the beam profile is smoother if the bubbles are smaller. This is because more MBs are necessary to achieve the same gas volume concentration, and the higher the number of scatterers gives a smoother beam profile of the scattered field. Finally, as the incident wave propagates through every MB population, it undergoes attenuation and speed of sound variations, resulting on a shift in the resonance frequency of the MBs.
	
	\subsubsection{Scattered pressure field: Harmonics} \label{sec:Poly_Harmonics}
	In this section we look at the different harmonics of the excitation pulse that are present in the scattered pressure field. These are obtained by decomposing the scattered signal into specific frequency bands using an 4th order Butterworth filter. These frequency bands are (i) the fundamental F0 $[0.7, 1.3]\;\mathrm{MHz}$, (ii) the second harmonic 2H $[1.7, 2.3]\;\mathrm{MHz}$ and (iii) the third harmonic 3H $[2.7, 3.3]\;\mathrm{MHz}$, where the intervals define the cutoff frequencies of the applied filter.
	
	Figure~\ref{fig:Comp_Scattered_Pressure_Harm} shows the harmonic contributions of the scattered pressure field for each of the considered populations. In the fundamental (F0) frequency band (top row of Fig.~\ref{fig:Comp_Scattered_Pressure_Harm}), we observe that the strongest scattered field is generated by the type 2 resonant MB suspension with a peak amplitude of $-1.23\;\mathrm{dB}$. The type 1 population with the below-resonance oscillating MBs has the second highest peak pressure of $-2.61\;\mathrm{dB}$ as the resonance frequency is closer to the excitation frequency, in comparison to the other two remaining cases. A significant observation is that the scattered field from the type 4 polydisperse population has a peak amplitude of $-7.1\;\mathrm{dB}$ and is stronger than the case of type 3 above-resonance MBs, which have a peak pressure of $-8.74\;\mathrm{dB}$. This can be explained due to the presence of MBs with a resonance frequency around 1 MHz in the polydisperse suspension. 
	
	In the second harmonic (2H) frequency band (middle row of Fig.~\ref{fig:Comp_Scattered_Pressure_Harm}), we observe that the scattered field of the type 2 resonant MBs is still the highest of all the four distinct cases. The peak amplitude in this case is $-10.3\;\mathrm{dB}$. The peak pressure of the type 3 above-resonance oscillating MBs is $-12.09\;\mathrm{dB}$, which is larger than the respective value of $-15.82\;\mathrm{dB}$ of the type 1 population with the below-resonance oscillating MBs. This is explained due to the fact that the resonance frequency of the system of the former is closer to the 2H frequency band around 2 MHz. The type 4 polydisperse distribution shows the weakest peak pressure amplitude of $-18.89\;\mathrm{dB}$. Compared to the monodisperse populations, hardly any constructive interferences are observed below the polydisperse suspension, due to the varying phases of the oscillations that result from the different sizes of the contrast bubbles.
	
	Finally for the third harmonic (3H) frequency band (bottom row of Fig.~\ref{fig:Comp_Scattered_Pressure_Harm}), the type 3 below resonance MBs exhibit the strongest scattered pressure field with a peak amplitude of $-17.49\;\mathrm{dB}$, as their resonance frequency of $3.9\;\mathrm{MHz}$ is closer to the 3H frequency band. Still, the type 2 resonant MBs scatter the second highest pressure field with a peak amplitude of $-18.49\;\mathrm{dB}$. Inside the MB suspension, the type 4 polydisperse MBs give a peak pressure of $-22.66\;\mathrm{dB}$. This is stronger than the peak of the pressure field of the type 1 below-resonance oscillating MBs ($-25.91\;\mathrm{dB}$), because the smaller MBs with a resonance frequency close to $3\;\mathrm{MHz}$ add to the strong scattering of the larger MBs. Similar to 2H, the type 4 polydisperse MBs hardly yield constructive interference below the suspension, as is the case for the type 1, 2 and 3 monodisperse MBs. This observation predicts that the uniformity of the size distribution of a population might have an impact on the nonlinear imaging artifacts downstream the population.	
	
	The cumulative scattered pressure field is the addition of the signals emitted from all the MBs in the population taken into account their individual position and therefore all the phase delays. A simplified expression is to linearly project the behavior of a single MB to the behavior of a whole population of MBs. Thus, the simulated pressure fields of the populations show similar behavior with the projected response of the single MB in  Fig.~\ref{fig:Radius_Frequency}.
	\subsubsection{Total pressure field: Attenuation and speed of sound variations}
\begin{figure}[t!]
		\centering
		\includegraphics{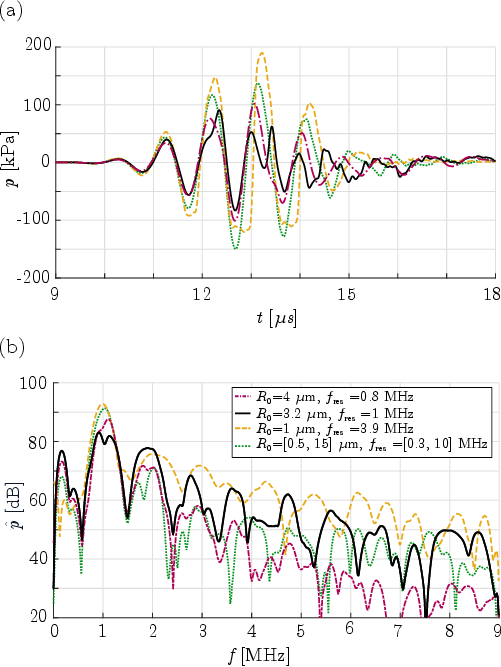}
		\caption{(a) Temporal signature and (b) frequency spectrum of the total pressure field, after propagation through each of the four distinct MB populations. The pressure is obtained for a point receiver located on the $z$-axis at a depth of $z=10.3\;\mathrm{mm}$.}		 
		\label{fig:TS_FullSpectrum}
	\end{figure}

	To show the influence of the nonlinear microbubble behavior on a propagating pressure wave, in Fig.~\ref{fig:TS_FullSpectrum} we show the temporal signatures and the respective frequency spectra after traversing each type of MB population. From Fig.~\ref{fig:TS_FullSpectrum}(a) it is clear that the type 2 monodisperse resonant population (black line) causes the most nonlinear distortion. The distortion takes place mainly after the second cycle as the MBs need to get a large oscillation amplitude before they demonstrate significant nonlinear behavior. The influence of the nonlinear bubble oscillation on the propagation through each one of the other three populations is much less visible in the time domain. By observing the frequency spectra in Fig.~\ref{fig:TS_FullSpectrum}(b), we can better see the effect of the nonlinear bubble behavior. Similar as in Sec.~\ref{sec:Poly_Harmonics}, the type 2 population of monodisperse oscillating MBs, shows a shift of energy from the fundamental to the second and higher harmonics. Furthermore, the maximum spectral amplitude of the fundamental is about equal for the other types of populations. The type 3 population of monodisperse below resonance MBs shows a strong second harmonic, and the highest third harmonic of all the populations, even higher than for the type 2 population.

\begin{figure}[t!]
		\centering
		\includegraphics{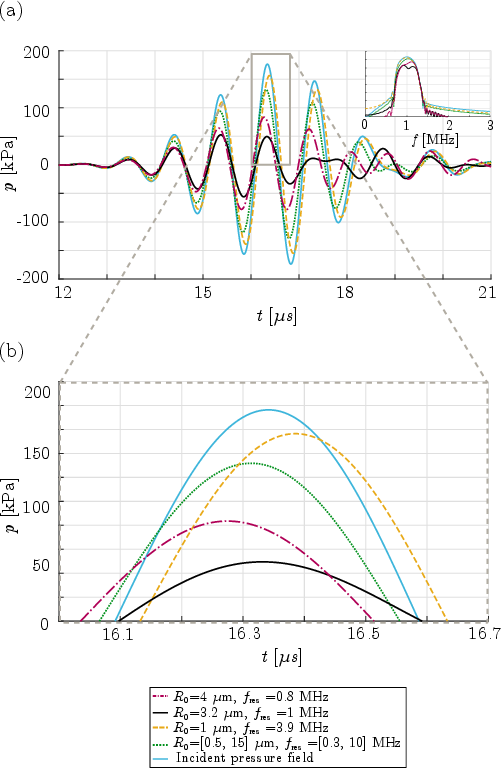}
		\caption{(a) Temporal signature of the total pressure field in the fundamental frequency band, after propagation through each of the four distinct MB populations. (b) Zoomed-in version of (a) to demonstrate the attenuation and speed of sound changes of the transmitted wave. The pressure is obtained for a point receiver located on the $z$-axis at a depth of $z=10.3\;\mathrm{mm}$.}		 
		\label{fig:TS_Fundamental}
	\end{figure}

	To quantify attenuation and speed of sound changes in the fundamental frequency band, we have subjected the temporal signatures in Fig.~\ref{fig:TS_FullSpectrum}(a) to a Butterworth filter of 8th order and a frequency pass band of $[0.75, 1.25]\;\mathrm{MHz}$. The results are plotted in Fig.~\ref{fig:TS_Fundamental}. For the type 1 population of bubbles that are below resonance, there is a decrease in peak pressure of $92.2\;\mathrm{kPa}$ relative to the incident field, and the speed of sound has been increased to $1517\;\mathrm{m/s}$. For the type 2 population with resonant bubbles, the peak pressure undergoes a drop of $126.9\;\mathrm{kPa}$, and the speed of sound has been maintained at $1482\;\mathrm{m/s}$. For the type 3 population of bubbles that are above resonance, the peak pressure experiences a drop of $19.9\;\mathrm{kPa}$, and the speed of sound has decreased to $1458\;\mathrm{m/s}$. Finally, for the type 4 polydisperse population, the decay in peak pressure is $44.8\;\mathrm{kPa}$, and the speed of sound has been increased to $1497\;\mathrm{m/s}$. We observe that the differences for the type 3 microbubbles are the smallest from all the populations, because they present the strongest effect mainly on the second harmonic. As in previous studies \cite{Sojahrood2023}, the INCS simulations demonstrate that for MBs with a resonance frequency below the excitation frequency there is an increase of the wave speed, whereas for a resonance higher than the excitation frequency there is a decrease of the wave speed. Finally, for the MBs with a resonance frequency equal to the excitation frequency, the wave speed is equal to the speed of sound of the background medium. 	%
	\subsubsection{Total pressure field: Convergence behavior}
	To quantify the numerical performance of our scheme, we  analyzed the difference between the successive iterations using the Relative Root Mean Square Error (RRMSE)
	\begin{equation}
		\mathrm{RRMSE} = \sqrt{\dfrac
			{\int_\mathcal{X_\mathrm{cd}}\int_\mathcal{T_\mathrm{cd}} {\left[p^{(j)}(\bx,t)-p^{(j-1)}(\bx,t)\right]^2}\,dt\, d\bx}
			{\int_\mathcal{X_\mathrm{cd}}\int_\mathcal{T_\mathrm{cd}} \left[p^{(0)}(\bx,t)\right]^2\,dt\, d\bx}},
		\label{eq:RRMSE}
	\end{equation}
	\begin{figure}[b!]
		\centering
		\includegraphics{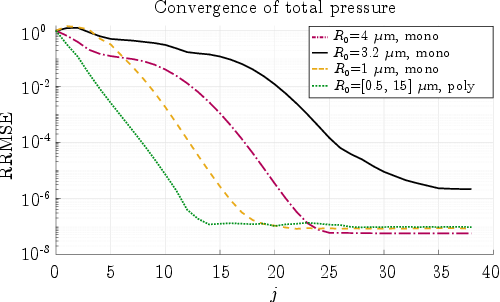}
		\caption{RRMSE as a function of the number of iterations $j$, for the considered types of MB populations.}		 
		\label{fig:Convergence}
	\end{figure}
	where $\mathcal{X_\mathrm{cd}}$ is the spatial computational domain, $\mathcal{T_\mathrm{cd}}$ is the temporal computational domain, $j$ is the iteration number and $p^{(j)}$ is the total pressure obtained in the $j$th iteration. The decay of the RRMSE is illustrated in Fig.~\ref{fig:Convergence} as a function of the number of iterations. A first observation is that after a certain number of successive iterations, the error tends to stabilize at a level of $10^{-5}$ or below. At this juncture, it can be inferred that incorporating additional multiple scattering orders will not yield further enhancements to the solution, indicating the attainment of insignificant scattering orders. Upon reaching this stage, it is assumed that the iterative process has converged to the lowest achievable error.
	
	For the type 2 monodisperse resonant MBs, it turns out that the initial iterations even show an RRMSE above 1. This indicates that the first multiple scattering orders are highly significant. Moreover, for these MBs more iterations are needed to reach convergence, and therefore more multiple-scattering orders should be included to achieve an accurate result. A general observation is that the closer the resonance frequency of the population is to the excitation frequency, the more iterations need to be taken into account. This can be explained by the fact that stronger close-range interactions occur in populations with resonant MBs due to the stronger scattering strength, making higher scattering orders more important. By observing the case of type 3 above-resonance monodisperse MBs, the RRMSE of the initial iterations is also above 1. This is due to the larger number of scatterers that are used to achieve the same gas volume concentration. This corresponds to higher number of bubble-bubble interactions at short distances. Finally, the type 4 polydisperse MBs yield a faster convergence (in the 13th iteration) than every other type of monodisperse suspension, demonstrating the relative significance of multiple scattering for the monodisperse populations.	
	%
	%================================================================================
	\subsection{CEUS imaging}
	\label{sec:MedicalApp}
	%================================================================================
	%
	%
	\begin{figure}[b!]
		\centering
		\includegraphics[width=8cm]{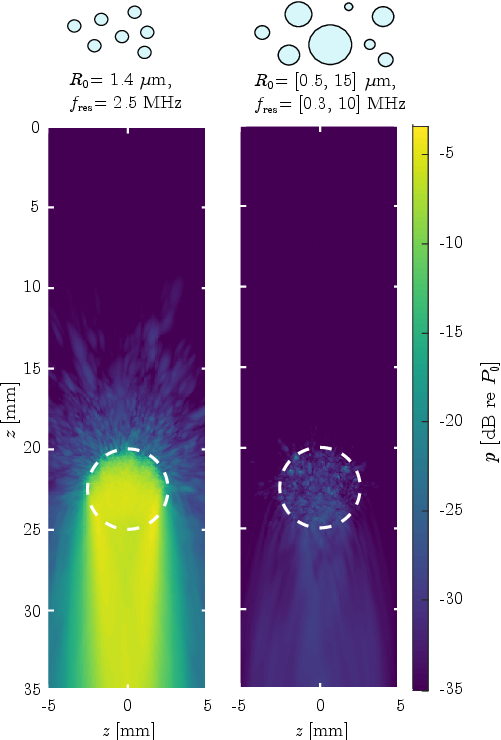}
		\caption{Residual acoustic pressure fields after the AM operation in the presence of (a) a type 5 resonant monodisperse population and (b) a type 6 polydisperse population. The MBs are located inside the dashed white circle.}		 
		\label{fig:ScatteredPressureFields}
	\end{figure}
	\subsubsection{Scattered pressure fields}\label{subsec:MA_P_sc}
	\begin{figure}[b!]
		\centering
		\includegraphics{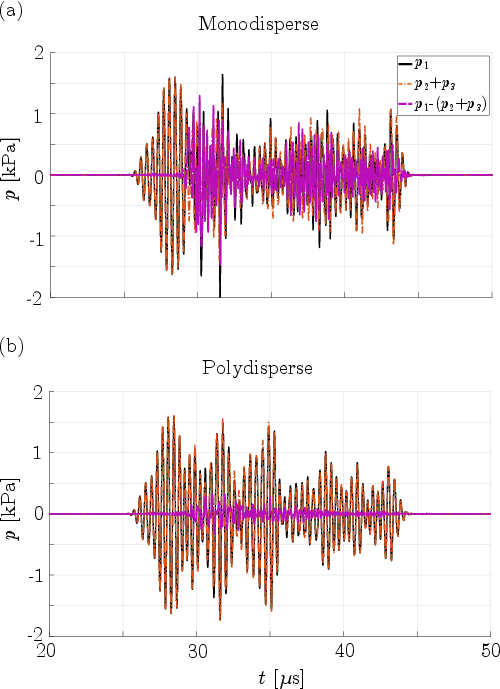}
		\caption{Temporal signatures used in the AM procedure for (a) a type 5 monodisperse resonant population and (b) a type 6 polydisperse population, both surrounded by tissue-mimicking LSs and measured at the center of the aperture of the linear array. Each of these graphs encompasses three plots, representing the double amplitude signal $p_1$,  the sum of the two respective single amplitude signals $p_2+p_3$, and the AM residual signal $p_1-(p_2+p_3)$.}		 
		\label{fig:TS_poly_vs_mono}
	\end{figure}
	In this section we compare the nonlinear scattering coming from suspensions of type 5 resonant monodisperse MBs and type 6 polydisperse MBs when these are surrounded by linear scatterers, as illustrated in Fig.~\ref{fig:ContrastDomain}(b). To resemble an in vivo setting and match the gas volume concentration, for the type 5 suspension, we use a concentration of $5\times 10^5\;\mathrm{ml}^{-1}$ MBs with $1.4\;\mu\mathrm{m}$ rest radius, corresponding to a total gas volume of $5.8\times10^{-6}\;\mathrm{ml}$. Furthermore, for the type 6, we use $3.1\times 10^4\;\mathrm{ml}^{-1}$ MBs, corresponding to the same total gas volume. First, the total pressure fields in these configurations are computed for three different excitations: field $p_1$ is due to a double amplitude excitation (full aperture), and the fields $p_2$ and $p_3$ result from two single amplitude excitations (odd elements and even elements), respectively. After employing the AM procedure, the peak residual AM pressures are as shown in Fig.~\ref{fig:ScatteredPressureFields}. For the monodisperse case in Fig.~\ref{fig:ScatteredPressureFields}(a), nonlinear effects accumulate in the suspension and propagate in the area below the population. The peak AM residual pressure is $-3.7\;\mathrm{dB}$ relative to the pressure at the source surface $P_0$. On the other hand, for the polydisperse case in Fig.~\ref{fig:ScatteredPressureFields}(b), the residual pressure field shows a relative peak amplitude of $-19.9\;\mathrm{dB}$, which is 6.5 times smaller than the respective for the monodisperse suspension. Most MBs in the polydisperse suspension are less efficient scatterers than the MBs in the resonant monodisperse population and, more importantly, bubbles with different sizes will cause nonlinear scattering with different phases. Therefore, the nonlinearities due to scattering do not propagate outside the MB domain. These results indicate that in CEUS the nonlinear wave propagation artifacts will be stronger for a resonant monodisperse population than a polydisperse population. 
	
	To demonstrate what this means for the AM imaging process, in Fig.~\ref{fig:TS_poly_vs_mono} we compare the time signatures of the double amplitude pulse $p_1$, the sum $p_2+p_3$ of the two single amplitude pulses, and the AM residual $p_1 - (p_2+p_3)$, for both the type 5 monodisperse and the type 6 polydisperse case. We depict the temporal signatures for the center of the aperture of the linear array. In Fig.~\ref{fig:TS_poly_vs_mono}(a), the AM residual of the monodisperse population is a strong signal with a peak pressure of \textbf{$1.5\;\mathrm{kPa}$}, compared to \textbf{$2.11\;\mathrm{kPa}$} for the incident double excitation field. The sum of the two single amplitude signals matches the waveform of the double amplitude signal only in the beginning of the pulse, which corresponds to the scattering of the LSs that are present above the MB suspension. The AM residual signal is stronger at the end of the pulses, which denotes the propagation of the nonlinear scattering of the MBs to the LSs that are located below the MB suspension.  
	
	In contrast, Fig.~\ref{fig:TS_poly_vs_mono}(b) shows that for the polydisperse case, the peak pressure of the AM residual corresponds to \textbf{$0.35\;\mathrm{kPa}$}, which is 4.3 times smaller than the respective value of the type 5 monodisperse population. Moreover, the sum of the single amplitude signals overlaps with the double amplitude signal, both in the beginning (scattering from the LSs above the MB suspension) and in the end (scattering from the LSs below the MB suspension) of the pulses. This indicates that the nonlinear scattering that propagates below the polydisperse MB suspension is relatively small.
	\subsubsection{Effect of size distribution on imaging artifacts}\label{sec:MA_Imaging}
	\begin{figure}[t!]
		\includegraphics{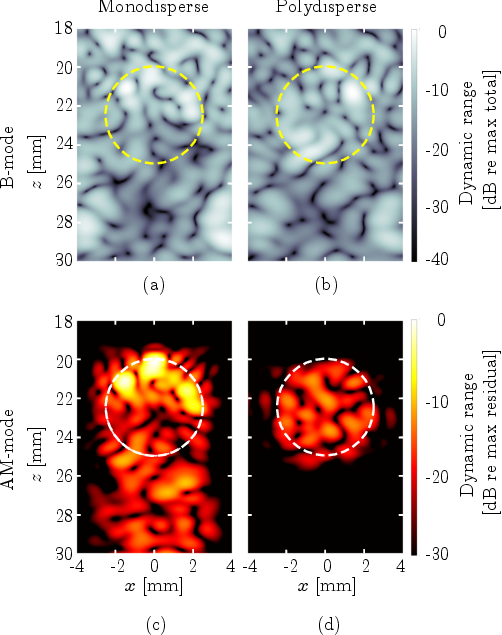}
		\caption{B-mode and AM-mode images of the population of monodisperse MBs and its surrounding region. B-mode (single-shot) ultrasound images acquired for a region with (a) type 5 monodisperse resonant population and (b) type 6 polydisperse MB population. AM ultrasound images acquired for the same regions of (c) monodisperse and (d) polydisperse MB populations. The position of the MB populations is outlined by a dashed circle. The rest of the simulation domain is filled with tissue-mimicking LSs.}		 
		\label{fig:BeamformedImg}
	\end{figure}
	To assess the imaging effects of the nonlinear fields  below each MB population, it is necessary to generate the reconstructed B-mode (single-shot) images and the images that are obtained after employing the AM procedure. The results are depicted in Fig.~\ref{fig:BeamformedImg}. To achieve this, we placed  $7\times 10^5\;\mathrm{ml}^{-1}$ tissue-mimicking linear scatterers (grey) surrounding the MB suspension. 
	
	Figures~\ref{fig:BeamformedImg}(a) and (b) depict the B-mode images for the configuration with a resonant monodisperse MB population and a polydisperse population, respectively. In both cases the backscattering from tissue-mimicking LSs and the MBs is indistinguishable because the areas with LSs and MBs have a similar echogenicity, independent of the size distribution. This demonstrates that B-mode imaging does not allow to disentangle nonlinear MB scattering from tissue mimicking scattering.
	
	Figures~\ref{fig:BeamformedImg}(c) and (d), show the AM images for the configuration with a resonant monodisperse MB population and a polydisperse population, respectively. Employing the AM sequence for imaging a monodisperse MB population generates an image with significant nonlinear artifacts below of the MB area, meaning that tissue scatterers get misclassified as MBs. On the contrary, applying the AM sequence for imaging the polydisperse population delivers an image with much higher specificity. The peak amplitude in the image of the monodisperse area ($0\;\mathrm{dB}$) is stronger than in the image of the polydisperse area ($-11.8\;\mathrm{dB}$). The peak value of the nonlinear artifact level is $-10.04\;\mathrm{dB}$ for the monodisperse population and $-29.4\;\mathrm{dB}$ for the polydisperse population. This is an indication that monodisperse MBs are more efficient scatterers than polydisperse populations, especially in applications that require to enhance deep tissue imaging. A drawback of CEUS with monodisperse MBs is that the generated artifacts due to propagation of nonlinear scattering in the area below the MBs, is of comparable magnitude and can lead to missclassification of tissue as contrast agents.
	%========================================================
	%=================================================================================================
	\section{Conclusions}\label{sec:Poly_Conclusions}
	%=================================================================================================
	We simulated AM ultrasound imaging of both monodisperse and polydisperse MBs using the INCS method, taking into account all the relevant physical phenomena occurring during ultrasound propagation through a MB population. We highlighted the significance of multiple scattering in monodisperse populations. Resonant monodisperse MBs are shown to be the most efficient scatterers, which corresponds to high sensitivity for CEUS. This property is crucial for optimizing contrast enhancement, guaranteeing consistent behavior and reliable imaging outcomes, especially compared to using polydisperse contrast agents. The drawback of resonant monodisperse MBs is the generation of imaging artifacts, which reduce the specificity of CEUS. This research approach is useful for optimizing CEUS imaging by designing the size distribution and parameters of a MB population through simulations.
	\section*{Acknowledgments}
	This research was supported by the project "Optoacoustic sensor and ultrasonic MBs for dosimetry in proton therapy" of the Dutch National Research Agenda, which is partly financed by the Dutch Research Council (NWO). The authors thank N. de Jong for his involvement in this research.
	
	%=======================================================
	
\end{document}